\documentclass[letterpaper,final]{iopart}

\usepackage{iopams,graphicx,times} 
 
\begin{document}

\letter{Surface critical behaviour at $\boldsymbol{m}$-axial Lifshitz
  points: continuum models, boundary conditions and two-loop
  renormalization group results}

\author{H.~W. Diehl, S.\ Rutkevich\footnote{On leave from: Institute
    of Solid State and Semiconductor Physics, 220072 Minsk, Belarus}
  and A.\ Gerwinski}

\address{Fachbereich Physik, Universit{\"a}t
Duisburg-Essen, D-45117 Essen, Germany}

\begin{abstract}
  The critical behaviour of semi-infinite $d$-dimensional systems with
  short-range interactions and an $O(n)$ invariant Hamiltonian is
  investigated at an $m$-axial Lifshitz point with an isotropic
  wave-vector instability in an $m$-dimensional subspace of
  $\mathbb{R}^d$ parallel to the surface.  Continuum $|\bphi|^4$
  models representing the associated universality classes of surface
  critical behaviour are constructed.  In the boundary parts of their
  Hamiltonians quadratic derivative terms (involving a dimensionless
  coupling constant $\lambda$) must be included in addition to the
  familiar ones $\propto\phi^2$. Beyond one-loop order the
  infrared-stable fixed points describing the ordinary, special and
  extraordinary transitions in $d=4+\frac{m}{2}-\epsilon$ dimensions
  (with $\epsilon>0$) are located at
  $\lambda=\lambda^*=\Or(\epsilon)$. At second order in $\epsilon$,
  the surface critical exponents of both the ordinary and the special
  transitions start to deviate from their $m=0$ analogues. Results to
  order $\epsilon^2$ are presented for the surface critical exponent
  $\beta_1^{\rm ord}$ of the ordinary transition. The scaling
  dimension of the surface energy density is shown to be given exactly
  by $d+m\,(\theta-1)$, where $\theta=\nu_{l4}/\nu_{l2}$ is the bulk
  anisotropy exponent.
\end{abstract}
%
\nosections
Lifshitz points, i.e.\ multicritical points at which a disordered, a
homogeneous ordered, and a modulated ordered phase meet, have been
known since the end of the 1970s [1--4]. 
Appropriate $n$-vector $|\bphi|^4$-models representing universality
classes of $m$-axial Lifshitz points were introduced at the same time;
the simplest ones have a Hamiltonian ${\mathcal H}=\int
d^dx\,{\mathcal L}_{\rm b}(\boldsymbol{x})$ with density
\begin{equation}\label{eq:Ham}
{\mathcal L}_{\rm b}=
\frac{\mathring{\sigma}}{2}\,{(\triangle_\alpha
 \bphi )}^2
+\frac{1}{2}\,{(\nabla_\beta\bphi )}^2
+\frac{\mathring{\rho}}{2}\,{({\nabla_\alpha}\bphi )}^2
+\frac{\mathring{\tau}}{2}\,
\bphi^2+\frac{\mathring{u}}{4!}\,|\bphi |^4\;,
\end{equation}
where the position vector
$\boldsymbol{x}\equiv(\boldsymbol{x}_\alpha,\boldsymbol{x}_\beta)$ has
$m$- and $(d-m)$-dimensional components $\boldsymbol{x}_\alpha$ and
$\boldsymbol{x}_\beta$, respectively, $\nabla_\alpha$ and
$\nabla_\beta$ denote the corresponding gradients and
$\triangle_\alpha$ means the Laplacian $\nabla_\alpha^2$.

Although the possibility of studying the universality classes of these
models in a systematic manner by means of expansions in $d$ and $m$
about general points on the line of upper critical dimensions
$d^*(m)=4+m/2$ ($0\le m\le 8$) had been realized already in 1975
\cite{HLS75a}, the enormous technical difficulties one encounters
beyond one-loop order [5--8] 
had prevented a successful implementation of this programme until
recently when a full two-loop renormalization group (RG) analysis was
performed and the $\epsilon=d^*(m)-d$ expansions of all critical
exponents were determined to order $\epsilon^2$ [9--13].

Here we are concerned with the effects of \emph{surfaces} on the
critical behaviour at such $m$-axial Lifshitz points. The only
previous studies of this problem we are aware of are restricted to the
uniaxial ($m=1$) Ising ($n=1$) case and use either the mean-field
approximation [13--15] 
or Monte Carlo simulations \cite{Ple02} for the ANNNI model. Let us
consider semi-infinite systems with a boundary plane $\mathfrak{B}$ at
$z=0$, where $z\ge 0$ is the Cartesian coordinate along the inner
normal on $\mathfrak{B}$.  Since $\boldsymbol{x}_\alpha$ and
$\boldsymbol{x}_\beta$ scale differently, two distinct basic
orientations of the surface plane exist which we call parallel and
perpendicular, depending on whether $\boldsymbol{n}$ is orthogonal to
the $\alpha$ or the $\beta$ subspace. We restrict ourselves here to
the case of parallel surface orientation; the case of perpendicular
orientation requires separate considerations and a distinct analysis
\cite{DGRunpub}.

For semi-infinite $|\bphi|^4$ models with an $O(n)$ symmetric
Hamiltonian three distinct types of surface transitions occurring at
the bulk critical point can be distinguished \cite{Die86a,Die97}: the
ordinary, special and extraordinary transitions.%
\footnote{The occurrence of the extraordinary and special transitions
  requires that the surface dimension $d-1$ is sufficiently high so
  that long-range surface order is possible in the presence of a
  disordered bulk.}
Analogues of these surface transitions should exist also for the
$m$-axial bulk Lifshitz points described by Hamiltonians with the bulk
density (\ref{eq:Ham}) (see footnote \footnotemark[2]). For the uniaxial Ising case
$m=n=1$ in $d=3$ dimensions, Pleimling's Monte Carlo results
\cite{Ple02} and the mean field analysis of \cite{FKB00}
lend support to this expectation. Our goal is to pave the ground for
systematic field theory analyses of these transitions.

To this end we need an appropriate semi-infinite extension of the bulk
model with the density (\ref{eq:Ham}). For the short-range interaction
case we are concerned with, it is justified to choose a Hamiltonian of
the form (with ${\mathbb R}^d_+\equiv {\mathbb R}^{d-1}\times
[0,\infty)$)
\begin{equation}
  \label{eq:Hamsi}
  {\mathcal H}=\int_{{\mathfrak V}={\mathbb R}^d_+}{\mathcal L}_{\rm
    b}(\boldsymbol{x})\,\rmd V
  +\int_{\mathfrak{B}}{\mathcal L}_1(\boldsymbol{x})\,\rmd A\;,
\end{equation}
where ${\mathcal L}_1(\boldsymbol{x})$ depends on
$\bphi(\boldsymbol{x})$ and its derivatives. We must now (i) find out
which contributions have to be retained in ${\mathcal L}_1$, (ii)
determine the boundary conditions they imply, (iii) clarify the
renormalization of the field theory  and set up a RG approach in
$d^*(m)-\epsilon$ dimensions, and (iv) derive the fixed-point
structure, identifying potential fixed points describing the ordinary,
special and extraordinary transitions.

In the case of a critical point (corresponding to the choice $m=0$),
it is sufficient to include a term $\propto \bphi^2$ in ${\mathcal
  L}_1$; other $O(n)$ (or ${\mathbb Z_2}$) invariant contributions can
be shown to be redundant or irrelevant \cite{Die86a,Die97}. However,
in the ($m\ne 0$) case of a Lifshitz point, this is not sufficient; we
must take
\begin{equation}
  \label{eq:L1}
  {\mathcal L}_1(\boldsymbol{x})=\frac{\mathring{c}}{2}\,\bphi^2
  +\frac{\mathring{\lambda}}{2}\,{\left(\nabla_\alpha\bphi\right)}^2\;.
\end{equation}
Power counting tells us that
$\mathring{\lambda}\,\mathring{\sigma}^{-1/2}$ is dimensionless.
Hence it is scale invariant at the Gaussian fixed point and
potentially infrared relevant for $\epsilon>0$. All other
contributions, notably terms
$\propto{\left(\nabla_\beta\bphi\right)}^2$,
$\propto\bphi\nabla_\alpha\bphi$ or $\propto\bphi\nabla_\beta\bphi$,
can be ruled out by symmetry or shown to be irrelevant or redundant
\cite{DGRunpub}. The field theory defined by equations
(\ref{eq:Ham})--(\ref{eq:L1}) satisfies the boundary conditions (valid
in an operator sense \cite{Die86a,Die97})
\begin{equation}
  \label{eq:bc}
  \partial_n\bphi=
  (\mathring{c}-\mathring{\lambda}\,\triangle_\alpha)\bphi\;.
\end{equation}
This carries over to the free propagator
$G(\boldsymbol{x},\boldsymbol{x}')$, whose Fourier transform,
$\hat{G}$, with respect to the $d-1$ coordinates parallel to the
surface reads, in the disordered phase,
\begin{equation}
  \label{eq:Gfree}
 \hat{G}(\boldsymbol{p};z,z')
=\frac{1}{2\kappa_{\boldsymbol{p}}}
  {\bigg[}\rme^{-\kappa_{\boldsymbol{p}}|z-z'|} 
-\frac{\mathring{c}+\mathring{\lambda}\,
  |\boldsymbol{p}_\alpha|^2-\kappa_{\boldsymbol{p}}}{\mathring{c}+
  \mathring{\lambda}\,|\boldsymbol{p}_\alpha|^2+
  \kappa_{\boldsymbol{p}}}\,\rme^{-\kappa_{\boldsymbol{p}}(z+z')}{\bigg]}
\end{equation}
with
\begin{equation}
  \label{eq:kappap}
  \kappa_{\boldsymbol{p}}=\sqrt{\mathring{\tau}+\mathring{\rho}\,
    |\boldsymbol{p}_\alpha|^2+|\boldsymbol{p}_\beta|^2+
    \mathring{\sigma}\, |\boldsymbol{p}_\alpha|^4}\;,
\end{equation}
where $\boldsymbol{p}_\alpha$ is the $m$-dimensional $\alpha$
component of the wave-vector $\boldsymbol{p}\in{\mathbb R}^{d-1}$. The
back transform of the part depending on $|z-z'|$ is the free bulk
propagator $G_{\rm b}(\boldsymbol{x}-\boldsymbol{x}')$. At the Gaussian
Lifshitz point $\mathring{\tau}=\mathring{\rho}=\mathring{u}=0$, it
takes the scaling form
\begin{equation}
  \label{eq:GbLP}
  G_{\rm b}(\boldsymbol{x})={|\boldsymbol{x}_\beta|}^{-2+\epsilon}\,
  \mathring{\sigma}^{-m/4}\,\Phi_{m,d}{\Big(\mathring{\sigma}^{-1/4}\,
    {|\boldsymbol{x}_\alpha|}\,{|\boldsymbol{x}_\beta|}^{-1/2}\Big)}\;.
\end{equation}
Here the scaling function $\Phi_{m,d}(\upsilon)$ is a generalization
of a generalized hypergeometric function (a Fox-Wright $_{1\!}\psi_1$
function, cf equations (10)--(13) of \cite{SD01}). In the special
cases $\mathring{c}= \mathring{\lambda} = 0$ or $\mathring{c}\to
\infty$ at arbitrary $\mathring{\lambda}\ge 0$,
$G(\boldsymbol{x},\boldsymbol{x}')$ reduces to the Neumann or
Dirichlet propagator, respectively, whose $z+z'$ dependent parts
reduce to $\pm G_{\rm
  b}(\boldsymbol{x}-\boldsymbol{x}'+2z'\,\boldsymbol{n})$.

Utilizing these results, and employing dimensional regularization in
conjunction with minimal subtraction of poles, we have performed a
two-loop RG analysis of the model (1)--(3) in $d^*(m)-\epsilon$
dimension.

Its main results are as follows. To renormalize the multi-point
correlation functions $G^{(N,M)}$ involving $N$ fields $\bphi$ off and
$M$ fields
$\bphi^{\mathfrak{B}}\equiv\bphi{(\boldsymbol{x}\in\mathfrak{B})}$ on
the boundary, the `bulk' re-parameterizations known from
\cite{DS00a,SD01},
\begin{eqnarray}
  \label{eq:bulkrep}
  \bphi=Z_\phi^{1/2}\,\bphi_{\rm ren}\;,\quad
  \mathring{\sigma}=Z_\sigma\,\sigma\;,
\quad
\mathring{\tau}-\mathring{\tau}_{\rm LP}=\mu^2\,Z_\tau\,\tau\;,
\nonumber\\ 
\left(\mathring{\rho}-\mathring{\rho}_{\rm LP}\right)\,
{\mathring{\sigma}}^{-1/2}=\mu\,Z_\rho\,\rho\;,\quad 
   \mathring{u}\,{\mathring{\sigma}}^{-m/4}\,F_{m,\epsilon}=
   \mu^\epsilon\,Z_u\,u\;,
\end{eqnarray}
must be complemented by `surface' re-parameterizations of the form
\begin{equation}
  \label{eq:surfrep}
\fl
  \bphi^{\mathfrak{B}}=(Z_\phi
  Z_1)^{1/2}\,\bphi^{\mathfrak{B}}_{\rm ren}\;,
\quad
  \mathring{c}-\mathring{c}_{\rm sp}=\mu\,Z_c\,c\;,
\quad
\mathring{\lambda}\,\mathring{\sigma}^{-1/2}=\lambda+
P_\lambda(u,\lambda,\epsilon)\;,
\end{equation}
where the surface renormalization factors $Z_1$ and $Z_c$ depend on
$u$ \emph{and} $\lambda$. The function
\begin{equation}
  \label{eq:Plambda}
  P_\lambda(u,\lambda,\epsilon)=\sum_{i,j=1}^\infty
  P_{\lambda}^{(i,-j)}(\lambda)\,u^i\,\epsilon^{-j}
 = \sum_{i,j=1}^\infty\sum_{k=0}^\infty P_\lambda^{(i,-j;k)}\,
 u^i\,\epsilon^{-j}\,\lambda^k 
\end{equation}
does not vanish at $\lambda=0$; although the one-loop coefficient
$P_\lambda^{(1,-1)}(\lambda)$ vanishes at $\lambda=0$, the graph
\hspace{0.5ex}
\raisebox{-0.5em}{\includegraphics[width=5em]{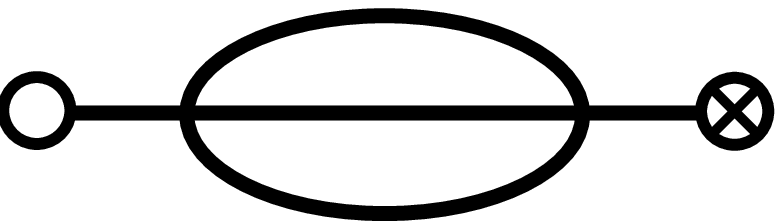}}
\hspace{0.5ex} of $\langle\phi\,\phi^{\mathfrak{B}}\rangle$ yields a
non-zero $P_\lambda^{(2,-1;0)}$. Thus a contribution $\propto
{\left(\nabla_\alpha\bphi\right)}^2$ to ${\mathcal L}_1$ gets
generated under the RG even if it was originally absent.

\begin{figure}[htbp]
  \centering
  \includegraphics[height=15em]{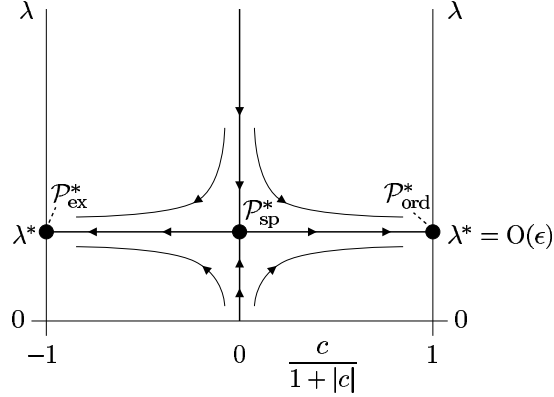}
  \caption{Schematic picture of the RG flow in the $c\lambda$
    plane if $m>0$, showing the fixed points ${\mathcal{P}}_{\rm
      ord}^*$, ${\mathcal{P}}_{\rm sp}^*$, and ${\mathcal{P}}_{\rm
      ex}^*$.}
  \label{fig:flow}
\end{figure}

The fixed points ${\mathcal{P}}_{\rm ord}^*$, ${\mathcal{P}}_{\rm
  sp}^*$ and ${\mathcal{P}}_{\rm ex}^*$ describing respectively the
ordinary, special, and extraordinary transitions must lie in the
$c\lambda$ plane at $(\tau,\rho,u)=(0,0,u^*)$, where $u^*$ is the
nontrivial root of the bulk beta function
$\beta_u(u)={\left.\mu\partial_\mu\right|_0}u$, computed to order
$\Or(\epsilon^2)$ in \cite{SD01}. For $u=u^*$, the beta function
$\beta_\lambda(u,\lambda)\equiv{\left.\mu\partial_\mu\right|_0}\lambda$
turns out to have an infrared-stable root at
\begin{equation}
  \label{eq:lambdastar}
  \lambda^*=-2\epsilon\,P_\lambda^{(2,-1;0)}/P_\lambda^{(1,-1;1)}
  +\Or(\epsilon^2)\;,
\end{equation}
with a correction-to-scaling exponent
\begin{equation}
  \label{eq:omegalam}\fl
  \omega_\lambda\equiv(\partial_\lambda\beta_\lambda)(u^*,\lambda^*) =
-P_\lambda^{(1,-1;1)}\,u^*+\Or(\epsilon^2)=
\frac{n+2}{n+8}\,\epsilon+\Or(\epsilon^2)\;.
\end{equation}
Thus ${\mathcal{P}}_{\rm ord}^*$, ${\mathcal{P}}_{\rm
  sp}^*$, and ${\mathcal{P}}_{\rm ex}^*$ are the fixed points at
$c=\infty$, $0$, and $-\infty$ displayed in figure \ref{fig:flow}.

Upon exploiting the RG equations implied by the above
re-parameterizations (\ref{eq:bulkrep}) and (\ref{eq:surfrep}) in a
standard fashion, one concludes that the critical surface exponents of
the special transition can be expressed in terms of bulk exponents and
\begin{equation}
  \label{eq:DelphiB}
  \Delta[\phi^{\mathfrak{B}}]=(d-m-2+\eta_{\rm L2}+\eta_1^{*,{\rm
      sp}}+m\theta)/2 = \beta_1^{\rm sp}/\nu_{\rm L2}
\end{equation}
and
\begin{equation}
  \label{eq:DelepsB}
  \Delta[\varepsilon^{\mathfrak{B}}]=d-m-2+m\theta-\eta_c^{*,{\rm
      sp}} \;,
\end{equation}
the scaling exponents of $\bphi^{\mathfrak{B}}(\boldsymbol{x})$ and
the boundary energy density
$\varepsilon^{\mathfrak{B}}(\boldsymbol{x})=
[\bphi^{\mathfrak{B}}(\boldsymbol{x})]^2/2$, respectively, where the
superscript `$*,{\rm sp}$' means values taken at ${\mathcal{P}}_{\rm
  sp}^*$. The $\epsilon$ expansions of these exponents, like those of
the bulk exponents, turn out to be independent of $m$ to order
$\epsilon$, but can be shown to be $m$ dependent at $\Or(\epsilon^2)$
\cite{DGRunpub}.  Thus, $\eta_1^{*,{\rm
    sp}}=-\frac{n+2}{n+8}\,\epsilon+\Or(\epsilon^2)=\eta_c^{*,{\rm
    sp}}+\Or(\epsilon^2)$.%
\footnote{The implied $\Or(\epsilon)$ results $\beta^{\rm
    sp}_1=\frac{1}{2}-\frac{\epsilon}{4}+\Or(\epsilon^2)$ and
  $\Phi=\frac{1}{2}-\frac{n+2}{n+8}\,\frac{\epsilon}{4}+\Or(\epsilon^2)$
  of $\beta_1^{\rm sp}$ and the surface crossover exponent $\Phi$ may
  be gleaned from equations (3.156e,b) of \cite{Die86a}.}

These statements about the $m$-dependence apply equally well to the
surface critical exponents of the ordinary transition. To demonstrate
this via explicit $\Or(\epsilon^2)$ results, note that $\beta_1^{\rm
  ord}$ can be expressed quite generally in terms of standard bulk
exponents $\nu_{\rm L2}$, $\eta_{\rm L2}$ (or $\beta_{\rm L}$), $\theta$ and a
single additional anomalous dimension $\eta_{1,\infty}^*$ as
\begin{equation}
\fl
  \label{eq:beta1ord}
\beta_1^{\rm ord}=(\nu_{\rm L2}/2)(d-m+\eta_{\rm L2}+m\theta+
\eta^*_{1,\infty})=\beta_{\rm L}+\nu_{\rm L2}\,(1+\eta_{1,\infty}^*/2)\;. 
\end{equation}
Our two-loop result for $\eta_{1,\infty}^*$ is
\begin{equation}
  \label{eq:eta1inf}
\fl
  \eta_{1,\infty}^*=-\frac{n+2}{6}\,u^*{\Big\{1+
    u^*\Big[j_1(m)- 
    J_u(m)\Big]\Big\}}+\Or{\big[}{(u^*)}^3\big]\;.
\end{equation}
Here $u^*$ is the fixed-point value whose $\epsilon$ expansion to
$\Or(\epsilon^2)$ is given in equation (60) of \cite{SD01}
while $J_u(m)$, defined by equations (49) and (50) of that reference,
is one of the four single integrals ($j_\phi$, $j_\sigma$, $j_\rho$,
$J_u$) in terms of which the two-loop series coefficients of the bulk
exponents were written there [see its equations (43)--(45) and (50)].
Finally,
\begin{equation}
  \label{eq:j1m}\fl
  j_1(m)=
  \frac{2^{10+m}\pi^{6+3m/4}\,\Gamma(m/2)}{\Gamma(2-m/4)\,
    \Gamma^2(m/4)}\, {\int_0^\infty}\rmd \upsilon\,\upsilon^{m-5}\,
  \Phi_{m,d^*}(\upsilon) {\int_0^\upsilon}\rmd y\,y^3\,\Phi_{m,d^*}^2(y)
\end{equation}
is a similar new integral which can be reduced to a single one. (Upon
rewriting ${\int_0^\infty}\rmd\upsilon{\int_0^\upsilon}\rmd{y}$ as
$\int_0^\infty\rmd{y}{\int_y^\infty}\rmd\upsilon$, the latter $\upsilon$
integration can be performed analytically to obtain a combination of
hypergeometric functions.)

Combining these results yields
\begin{eqnarray}
\label{eq:eta1infeps}
\fl
\eta_{1,\infty}^*=-\frac{n+2}{n+8}\,\epsilon-\frac{n+2}{16(n+8)^3}
\bigg\{(n+2)\Big[ 
\frac{j_\sigma(m)}{m+2}-8\,j_\phi(m)\Big]+64\,(5n+22)\,J_u(m)\nonumber\\
\lo +96(8+n)\Big[j_1(m)- 
    J_u(m)\Big]\bigg\}\epsilon^2
+\Or(\epsilon^3)\;.
\end{eqnarray}

Just as for the above-metioned four integrals of \cite{SD01}, the
values of $j_1(m)$ can be computed analytically for the special
choices $m=2$, $m=6$ and $m\to 0$. This yields
\begin{equation}
  \label{eq:j1val}
  j_1(0)=\frac{1}{2}\;,\quad j_1(2)=1-\frac{\ln 3}{2}\;, \quad
  j_1(6)=-\frac{2}{3}+2\ln\frac{27}{16}\;.
\end{equation}
To determine $j_1(m)$ for other values of $m$ we had to resort to
numerical means of the kind utilized in \cite{SD01}. For the uniaxial
case $m=1$, we obtained $j_1(1)=0.47289(1)$. Note also that in the
limit $m\to 0$, the result (\ref{eq:eta1infeps}) reduces to the
familiar one for the standard semi-infinite $|\phi|^4$ model, given in
equation (IV.35) of \cite{DD81a}.

Let us briefly explain how the above results were obtained. Since the
fixed point ${\mathcal{P}}_{\rm ord}^*$ is located at $c=\infty$, the
ordinary transition can be investigated without having to retain the
full dependence on $c$ and $\lambda$. To see this, note that the free
propagator and the regularized bare $G^{(N,M)}$ become independent of
$\mathring{\lambda}$ in the limit $\mathring{c}\to\infty$, and
satisfy a Dirichlet boundary condition, which carries over to the
renormalized theory. The long-scale behaviour of the $G^{(N,M)}$ with
a nonzero number of $\bphi^{\mathfrak{B}}$ can be inferred from the
theory with $\mathring{c}=\infty$ and $\mathring{\lambda}=0$ via the
near-boundary behaviour of the operator $\phi$. To this end, one
considers correlation functions involving an arbitrary number of the
operators $\phi$ and $\partial_n\bphi$, and then uses the boundary
operator expansion (BOE) $\bphi_{\rm
  ren}(\boldsymbol{x}_{\mathfrak{B}}+z\boldsymbol{n})
\mathop{\approx}\limits_{z\to 0}C_{\rm ord}(z)\,\partial_n\bphi_{\rm
  ren}(\boldsymbol{x}_{\mathfrak{B}})$. The renormalized theory
requires in addition to the bulk re-parameterizations, the
multiplicative re-parameterization (\ref{eq:bulkrep}),
\begin{equation}
  \label{eq:normalderren}
  \partial_n\bphi=[Z_{1,\infty}(u)\,Z_\phi(u)]^{1/2}\,
  \partial_n\bphi_{\rm ren} 
\end{equation}
and an additive surface counter-term subtracting the primitive
divergence ($\propto p_\alpha^2$) of
$\langle\partial_n\bphi\,\partial_n\bphi\rangle$. The resulting RG
equations imply scaling and yield the behaviour $C_{\rm ord}(z)\sim
z^{1+\eta_{1,\infty}^*/2}$, where $\eta_{1,\infty}^*$ is the
fixed-point value of the exponent function associated with
$Z_{1,\infty}$. A straightforward consequence is that the exponents
characterizing the leading infrared singularities of the $G^{(N,M)}$
can be expressed in terms of (4 independent) bulk critical indices and
a single surface one, namely, $\eta_{1,\infty}^*$ or $\beta_1^{\rm
  ord}$. Upon computing $Z_{1,\infty}$ and its exponent function to
two-loop order and making extensive use of the results of 
\cite{DS00a} and \cite{SD01}, we arrived at equations
(\ref{eq:beta1ord})--(\ref{eq:eta1infeps}).

Let us also note that the scaling dimension of the surface energy
density at the ordinary fixed point is given exactly by
\begin{equation}
  \label{eq:Delbendens}
  \Delta^{\rm ord}[\varepsilon^{\mathfrak{B}}]=d+m\,(\theta-1)\;.
\end{equation}
That is , the leading thermal singularity of
$\varepsilon^{\mathfrak{B}}$ has the bulk-free energy form $\sim
|\tau|^{2-\alpha_L}$ with $\alpha_L=\nu_{\rm L2}(d-m +m\theta)$. The
result can be obtained in a variety of ways, namely: (i) by
generalizing the analysis given in appendix C of \cite{DDE83}, (ii) by
showing that the operator with smallest scaling dimension appearing in
the BOE of the energy density $\varepsilon(\boldsymbol{x})$ is the
component $T_{zz}$ of the stress-energy tensor [whose scaling
dimension is given by equation (\ref{eq:Delbendens})] and (iii) by
proceeding as in the derivation for the $m=0$ case given in section
III.B of \cite{BD94}.

The results (\ref{eq:eta1inf}) and (\ref{eq:eta1infeps}) can be
combined with known bulk results \cite{SD01} to estimate the values of
surface critical exponents such as $\beta_1^{\rm ord}$ for $d=3$. In
the uniaxial Ising case $m=n=1$, equation (\ref{eq:eta1infeps})
becomes
\begin{equation}
  \label{eq:eta1inf11}
  \eta_{1,\infty}^*(m=n=1) = - 0.333\overline{3}\,\epsilon 
  -0.1804\,\epsilon^2+\Or(\epsilon^3)\;,
\end{equation}
which gives $\eta_{1,\infty}^*\simeq-0.906$ if we set $\epsilon=3/2$
(i.e.\ $d=3$), truncating the series at order $\epsilon^2$. Inserting
this value together with the bulk estimates $\beta_{\rm L}\simeq
0.246$, $\nu_{\rm L2}\simeq 0.746$ and $\eta_{\rm L2}\simeq 0.124$ of
\cite{SD01} into equation (\ref{eq:beta1ord}) and the analogous
expressions $\gamma_{1}^{\rm ord} = \nu_{\rm L2}\,(1-\eta_{\rm L2}-
\eta_{1,\infty}^*/2)$ and $\gamma_{11}^{\rm ord} = -\nu_{\rm L2}\,
(1+\eta_{\rm L2} + \eta_{1,\infty}^*)$ for the surface susceptibility
exponents yields
\begin{equation}
  \label{eq:deq3est}
  \beta_1^{\rm ord}\simeq 0.65\;,\quad
  \gamma_{1}^{\rm ord}\simeq 0.99\;, \quad \gamma_{11}^{\rm
    ord}\simeq-0.2\quad(m=n=1,d=3)\;.
\end{equation}
Owing to the low order $\epsilon^2$ of the available series expansions
and the large value $\epsilon=3/2$ involved, these estimates cannot be
trusted to be very precise. (They inherit, in particular, any
uncertainty of the inserted bulk exponents.) However, they compare
reasonably well with Pleimling's recent Monte Carlo estimates
\cite{Ple02} $\beta_1^{\rm ord}=0.687(5)$, $\gamma_1^{\rm
  ord}=0.82(4)$ and $\gamma_{11}^{\rm ord}=-0.29(6)$.

In summary, we have identified the continuum models that represent the
universality classes of the considered ordinary, special and
extraordinary (surface) transitions at $m$-axial bulk Lifshitz points,
clarified their fixed point structure and presented two-loop RG
results. A more detailed account of this work will be presented
elsewhere \cite{DGRunpub}.

\ack
We gratefully acknowledge partial support by the Deutsche
Forschungsgemeinschaft (DFG) via Sonderforschungsbereich 237 and grant
Di-378/3.

\section*{References}

\end{document}